\documentclass[journal]{IEEEtran}

\usepackage{graphicx, amssymb,verbatim,amsmath,eepic,mathtools}

\newcommand{\ra}{\rightarrow}

\newcommand{\sR}{\mathbb R}
\newcommand{\sN}{\mathbb N}

\newcommand{\limes}[2]{\mathop {\lim }\limits_{#1 \to #2}}
\newcommand{\qlim}[1]{\limes{q}{#1}}

\newtheorem{theorem}{Theorem}
\newtheorem{remark}{Remark}
\newtheorem{lemma}{Lemma}

\begin{document}

\title{Comments on ``Generalization of Shannon-Khinchin axioms to nonextensive systems and the uniqueness theorem for the nonextensive entropy''}

\author{Velimir M. Ili\'c,
        Miomir S. Stankovi\'c, and
        Edin H. Mulali\'c
\thanks{V. Ili\'c is with the Mathematical Institute of the Serbian Academy of Sciences and Arts, Kneza Mihaila 36, 11000 Beograd, Serbia,}
\thanks{M. Stankovi\'c is with the Faculty of Occupational Safety, University of Ni\v s,
Serbia,}
\thanks{E. Mulali\'c is with the Mathematical Institute of the Serbian Academy of Sciences and Arts, Kneza Mihaila 36, 11000 Beograd, Serbia}
\thanks{This research was supported by the Ministry of Science and
Technological Development, Republic of Serbia, Grant No.
III44006}}

\maketitle


\begin{abstract}
Recently, Suyari has proposed a generalization of Shannon-Khinchin
axioms, which determines a class of entropies containing the well-known Tsalis and Havrda-Charvat entropies [\textit{H. Suyari, IEEE Trans. Inf. Theory, vol. 50, pp. 1783-1787, Aug. 2004}]. In
this comment we show that the class of entropy functions
determined by Suyari's axioms is wider than the one proposed by
Suyari and give a counterexample. Additionally, we generalize
Suyari's axioms characterizing recently introduced class of
entropies obtained by averaging pseudoadditive information content
introduced in [\textit{V. Ili\'c and M. Stankovi\'c,``Comments on ``Nonextensive Entropies derived
from Form Invariance of Pseudoadditivity'''' Submited, 2012}].
\end{abstract}



\section{Generalized Shannon-–Khinchin Axioms}

In this section we review generalized Shannon-Khinchin axioms. We
discuss the proof for unique class satisfying the axioms proposed
by Suyari \cite{suyari2004generalization} and give a counterexample.

\subsection{Suyari's theorem} \label{introduction}

Let $\Delta_n$ be the $n$-dimensional simplex,
\begin{equation}
   \Delta_n \equiv \left\{ (p_1, \dots , p_n) \Big\vert \; p_i \ge 0,
      \sum_{i=1}^{n} p_i = 1 \right \}
  \label{Delta}
\end{equation}
and let $\sR^+$ denotes the set of positive real numbers.

For a function $S_q: \Delta_n \ra \sR^+$, $q \in \sR^+$, $n \in \sN$, we
define the following Shannon-Khinchin (SK) axioms
[GSK1]$\sim$[GSK4]:


{[GSK1]} {\it continuity}:
         $S_q$ is continuous in $\Delta_n$ and with respect to $q \in \sR^+$;

{[GSK2]} {\it maximality}:
        for any $n \in N$ and any $(p_1, \dots, p_n) \in \Delta_n$
\begin{equation}
    S_q(p_1, \dots, p_n) \le S_q(\frac{1}{n},\dots ,\frac{1}{n});
\end{equation}

{[GSK3]} {\it expandability}:
      \begin{equation}
        S_1(p_1, \dots, p_n, 0) = S_1(p_1,\dots ,p_n);
      \end{equation}

{[GSK4]} {\it generalized Shannon additivity}:
   if
\begin{align}
  p_{ij} \ge 0, \quad &p_i \equiv \sum_{j=1}^{m_i} p_{ij}, \quad
  p(j \vert i) \equiv \frac{p_{ij}}{p_i}, \nonumber \\
  &\forall i=1,\dots, n, \quad
  \forall j=1, \dots, m_i,
\end{align}
 then the following equality holds:
\begin{align}
   \label{q-additivity}
    &S_q(p_{11}, \dots, p_{n m_n})\nonumber\\
    &=S_q(p_1,\dots ,p_n)
    + \sum_{i=1}^n p_i^q S_q \left( p(1 \vert i),\dots ,p(m_i \vert i) \right).
\end{align}


\begin{remark}

[GSK1]-[GSK4] reduces to [SK1]-[SK4] for $q=1$. As shown in \cite{khinchin1957mathematical}, $S_1$ is
Shannon entropy, i.e.
\begin{equation}
S_1\mathop = - k\sum\limits_{i=1}^n {p_i\ln p_i},
\label{Shannonentropy}
\end{equation}

where $k>0.$
Because of [GSK1], we have
\begin{equation}
\qlim{1}S_q=S_1\mathop = - k\sum\limits_{i=1}^n {p_i\ln p_i}.
\label{Shannonentropy_lim}
\end{equation}
\end{remark}

\begin{theorem}
Let $S_q: \Delta_n \ra \sR^+$, $q \in \sR^+$, $n \in \sN$ be a function which is not identically equals to zero for $n>1$ and
satisfies [GSK1]-[GSK3] and the following generalized Shannon
additivity axiom for $q \in \sR^+$. Then, $S_q: \Delta_n \ra
\sR^+$, $n \in \sN$, is uniquely determined with
\begin{equation}
\label{suyari: Sq}
    S_q(p_1, \dots, p_n) = \frac{1-\sum_{i=1}^{n} p_i^q}{ \phi(q)},
\end{equation}
for $q \in \sR^{+} \setminus \{1\}$ and $(p_1, \dots, p_n) \in
\Delta_n$, and $\phi(q)$ satisfies the following properties
i)-iv):
\begin{itemize}
  \item[i)] $\phi(q)$ is continuous and has the same sign as $q-1$;
  \item[ii)] $\lim_{q \to 1} \phi(q) = 0$, and $\phi(q) \ne 0$ for $q \ne 1$;
  \item[iii)] there exists an interval $(a, b) \in \sR^{+}$ such that $a<1<b$
   and $\phi(q)$ is differentiable on the interval $(a,1) \cup (1,b)$;
  \item[iv)] there exists a positive constant $k$ such that
              $\lim_{q \to 1} \frac{d \phi(q)}{d q} = \frac{1}{k}$.
\end{itemize}

\end{theorem}

According to Suyari's proof, the form (\ref{suyari: Sq}) is
uniquely determined by solving functional equation
(\ref{q-additivity}) from [GSK4] and by using the continuity with
respect to $p$ as assumed in [GSK1]. Property (i) follows from
maximality condition [GSK2]. The continuity with respect to $q$
from [GSK1] implies condition (ii). [GSK4] is used only for proof
of the basic case $q=1$, at which [GSK1]$\sim$[GSK4] reduce to
Shannon-Khinchin axioms.

To satisfy (\ref{Shannonentropy_lim}), Suyari, using the fact that numerator
${1-\sum\limits_{i=1}^n {p_i^q}}$ is differentiable with respect
to $q$, concludes that properties (iii) and (iv) must be satisfied
for $\phi \left( q \right)$. However, (\ref{Shannonentropy_lim}) will
actually be satisfied iff:
\begin{align}
\qlim{1} S_q(p_1&, \dots, p_n) =%
\qlim{1} {q - 1 \over \phi(q)}
\cdot \qlim{1} {1-\sum_{i=1}^{n} p_i^q \over q - 1}\nonumber\\
&=\qlim{1} {q - 1 \over \phi(q) - \phi(1)}
\cdot \qlim{1} {\left(-\sum_{i=1}^{n} p_i^q \ln p_i\right)}\nonumber\\
&=- k \cdot \sum_{i=1}^n p_i \ln p_i%
\end{align}
i.e. iff
\begin{equation}
\qlim{1} {\phi(q) - \phi(1) \over q - 1} = {1 \over k}
\end{equation}
(We have used L'Hospital's rule to show that $\qlim{1}
{(1-\sum_{i=1}^{n} p_i^q) / (q - 1)} = - \sum_{i=1}^n p_i \ln
p_i$.) Accordingly, properties (iii) and (iv) should be
replaced with
\begin{itemize}
  \item[iii$^\prime$)]
$\phi(q)$ is differentiable in $q=1$ and
\begin{equation}
\frac{d \phi(q)}{d q}\Big|_{q=1}={1 \over k}.
\end{equation}
\end{itemize}
Hence, the function $\phi(q)$ need not be differentiable in a
neighbourhood of $q=1$, but only in $q=1$. In the following subsection
we construct the function which has these two properties.

Condition (iii$^{\prime}$) describes a class of functions which are not characterized by conditions (iii)-(iv). One example is shown in the following subsection. Although Suyari's conditions (iii)-(iv) may appear as conditions which do not require differentiability of function $\phi(q)$ in 1 and therefore describe a class of functions not characterized by condition (iii$^\prime$), that is actually not the case, as the following Lemma shows.

\begin{lemma}
\label{lemma continuous derivative}
Let $f$ be a continuous function on $(a, b)$ and suppose that $x_0 \in (a, b)$ is such that $f^{\prime}(x)$ exists for all $x_0 \in (a, x_0) \cup (x_0, b)$. If $\lim_{x \to x_0}f^{\prime}(x) = L,$ then $f^{\prime}(x_0) = L.$
\end{lemma}
\begin{IEEEproof}
For sufficiently small $h \in \sR, h>0,$ $f$ is continuous on $[x_0, x_0+h]$ and $f^{\prime}(x)$ exists for all $x \in (x_0, x_0+h).$ Conditions for the Mean Value Theorem are fulfilled, therefore there exists $\eta_h \in (x_0, x_0+h)$ such that
\[
f(x_0 + h) - f(x_0) = h \cdot f^{\prime}(\eta_h).
\]
Hence
\begin{align*}
f_+^{\prime}(x_0) &= \lim_{h \to 0^+} \frac{f(x_0+h) - f(x_0)}{h}	\\
				  &= \lim_{h \to 0^+} \frac{h \cdot f^{\prime}(\eta_h)}{h}	\\
				  &= \lim_{h \to 0^+} f^{\prime}(\eta_h) = L.	\\	
\end{align*}
Similarly, it can be proven that $ f_-^{\prime}(x_0) = L, $ and therefore $f^{\prime}(x_0) = L. $  

\end{IEEEproof}

\subsection{Counterexample}

The Weierstrass function is a well known example of nowhere
differentiable continuous function \cite{hardy1916weierstrass}. It
is defined with:
\begin{equation}
W(x)=\sum_{k=0}^\infty a^k \cos\left(b^k \pi x \right),
\end{equation}
where $0<a<1$ , $b$ is a positive odd integer, $ab > 1 + 3\pi / 2$
and $x \in \sR$. The Weierstrass function is bounded, since
\begin{equation}
\label{counterexample: weierstrass bound} |W(x)| \leq
\sum_{k=0}^\infty a^k | \cos\left(b^k \pi x \right)| \leq
\sum_{k=0}^\infty a^k = W(0) < \infty,
\end{equation}
where $W(0)=1 / (1-a)$. 
Using the Weierstrass function we construct $\phi(q)$, which satisfies
properties (i), (ii) and (iii$^\prime$), but not properties (iii) and (iv).

Let
\begin{equation}
\label{counterexample: phi} \phi(q) = {q-1 \over k} \cdot { W(q -
1) + 2 \cdot W(0)  \over 3 \cdot W(0) }.
\end{equation}
Since $W(x)$ is continuous and $W(x) + 2W(0) > 0$ according to
(\ref{counterexample: weierstrass bound}), $\phi(q)$ satisfies
properties (i) and (ii). Moreover,
\begin{equation}
\frac{d \phi(q)}{d q}\Big|_{q=1}
= \qlim{1} {\phi(q) - \phi(1) \over q - 1} = %
 \qlim{1} {\phi(q) \over q - 1}={1\over k}
\end{equation}
and function $\phi(q)$ satisfies property (iii$^\prime$),

However, function $\phi(q)$ does not satisfy property (iii)
from Suyari's theorem since it is differentiable only in $q=1$.
Oppositely, the function
\begin{equation}
{1 \over q-1}\cdot \phi(q) = {1 \over k} \cdot { W(q - 1) + {2
\cdot W(0)}  \over {3 \cdot W(0)} }
\end{equation}
should be differentiable for some $q \neq 1$ as a product of
differentiable functions, further implying differentiability of
$W(q - 1)$, which is impossible since the Weierstrass function is
nowhere differentiable.

\section{New axiomatic system}

In this section we review the class entropies obtained as
averaging of pseudoadditive information content introduced in
\cite{ilic2012entropy}. After that we generalize Suyari's axiomatic system, which
uniquely determines the class of entropies derived in \cite{ilic2012entropy}.

\subsection{Entropy as expected information content}
\label{entropy as expected ic}

\begin{theorem}\rm
Let $I_q(p)$ be a function of two variables $q\in \sR^+$ and $p
\in \left( {0,1} \right]$, which satisfies the following axioms

{[S0]} $I_1\left( p \right)=- k \ln p,\quad k>0$,

{[S1]} $I_q$ is continuous with respect to $p\in\left(
{0,1}\right]$ and $q\in \sR^+$,

{[S2]} $I_q\left( {p} \right)$ is convex with respect to $p\in
\left(0,1\right]$ for any fixed $q\in \sR^+$,

{[S3]} There exists a function $\phi : \sR^+ \to \sR$ such that
\begin{equation*}
{I_q\left( {p_1p_2} \right)}={I_q\left( {p_1}\right)} +{I_q\left(
{p_2} \right)} +{\phi\left( q \right)}\cdot{I_q\left(
{p_1}\right)}\cdot
{I_q\left( {p_2} \right)} 
\end{equation*}
for any $p_1,p_2\in \left( {0,1} \right]$, $\phi(q) \neq 0$ for $q
\neq 1$ and $\phi(q)$ is continuous \footnote{The axiom [S3] from \cite{ilic2012entropy} is actually not the same [S3] from the present paper, but it is  equivalent. It can be obtained if we set $\phi(q) = \varphi(q) / k$, where $\varphi: \sR^+ \ra \sR$ is continuous. }.

Then, the unique nontrivial solution is given by
\begin{equation}
\label{information content}
I_q\left( p \right)=%
{1 \over {\phi \left( q \right)}}%
\cdot \left( p^{\alpha(q)} -1 \right),
\end{equation}
where $k$ is a positive constant and
\begin{description}

\item{(a)} $\alpha \left( q \right)$ is continuous with
respect to any $q\in \sR^+$, $\alpha(1)=0$, $\alpha(q)\neq 0$ for $q \neq 1$ and

\begin{equation}
\qlim{1} {\alpha \left( q \right) \over \phi \left( q \right)} =
-k.
\end{equation}

\item{(b)} it holds that
\begin{equation}
\alpha(q) \in
\begin{dcases}
(-\infty,0] 
\;\; &\hbox{for} \;\; \phi(q) > 0 \\
\quad
[0 , 1] \;\; &\hbox{for} \;\; \phi(q) < 0.
\end{dcases}
\end{equation}

\end{description}
\end{theorem}

\begin{remark}
Note that condition $\alpha(q)\neq 0$ for $q \neq 1$ ensures that information content is not identically equals zero for some $q$.
\end{remark}

Nonextensive entropy of distribution $p$, $S_q\left( p \right)$,
is defined as the appropriate expectation value of 
$I_q$,
\begin{equation}
\label{gexp: nonextensive entropy}%
S_q\left(p \right)\equiv %
E_{q,p}\left[ {I_q} \right].
\end{equation}
The expectation is chosen so that the maximality principle is satisfied:
\begin{equation}
    S_q(p_1, \dots, p_n) \le S_q(\frac{1}{n},\dots ,\frac{1}{n}).
\end{equation}
The simplest case is the trace form expectation:
\begin{equation*}
E_{q,p}\left[ I_q \right]\equiv\sum\limits_{i=1}^n  e_q(p_i) \cdot I_q(p_i)
\end{equation*}
In this case the maximality condition is satisfied if $e_q(p) \cdot
I_q(p)$ is concave as shown in \cite{hanel2011comprehensive}. For the
information content (\ref{information content}) one possible
choice is $e_q(p) = p^{-\alpha(q)+1}$ in which case we obtain
\begin{equation}
\label{invar: S unnor}%
S_q \left( p \right) =%
{1-\sum\limits_{i=1}^n {p_i^{-\alpha(q) +1}}  \over
\phi(q) }.
\end{equation}
Note that the trace form expectation operator (\ref{gexp:
nonextensive entropy}) represents the generalization of
expectation operator which is used in axiom [GSK4] for
$\alpha(q)=1-q$. In the following subsection we will give the
axiomatization of entropy (\ref{invar: S unnor}) by generalization
of the axiomatic system [GSK1]$\sim$[GSK4] from section
\ref{introduction} based on expectation operator (\ref{gexp:
nonextensive entropy}).

\subsection{New axiomatic system}

\begin{theorem}

Let $S_q: \Delta_n \ra \sR^+$, $q \in \sR^+$, $n \in \sN$ be a function which is not identically equals to zero for $n>1$ and
satisfies [GSK1]-[GSK3] and the following generalized Shannon
additivity axiom for $q \in \sR^+$.

{[gsk4]} {\it generalized Shannon additivity}:
   if
\begin{align}
  p_{ij} \ge 0, \quad &p_i \equiv \sum_{j=1}^{m_i} p_{ij}, \quad
  p(j \vert i) \equiv \frac{p_{ij}}{p_i}, \nonumber \\
  &\forall i=1,\dots, n, \quad
  \forall j=1, \dots, m_i
\end{align}
and $\alpha: \sR^+ \ra \sR$ is a continuous function, then the following equality holds:
\begin{align}
   \label{q-additivity_g}
    &S_q(p_{11}, \dots, p_{n m_n})\nonumber\\
    &=S_q(p_1,\dots ,p_n)
    + \sum_{i=1}^n p_i^{-\alpha(q)+1} S_q \left( p(1 \vert i),\dots ,p(m_i \vert i) \right),
\end{align}
where $\alpha(q)$ is continuous, $\alpha(1)=0$ and $\alpha(q)\neq 0$ for $q \neq 1$.

Then, $S_q: \Delta_n \ra \sR^+$, $n \in \sN$, is uniquely
determined with
\begin{equation}
\label{suyari: Sq_g}
    S_q(p_1, \dots, p_n) =\frac{1-\sum_{i=1}^{n} p_i^{-\alpha(q)+1}}{ \phi(q)},
\end{equation}
where  $(p_1, \dots, p_n) \in \Delta_n$, $q \in \sR^{+}$ and

\begin{description}

\item{(a)} $\phi \left( q \right)$ is continuous with
respect to any $q\in \sR^+$, $\phi(q) \ne 0$ for $q \ne 1$ and $\phi(1)=0$ [OP] and

\begin{equation}
\label{new ax: alpa phi limit} \qlim{1} {\alpha \left( q \right)
\over \phi \left( q \right)} = -k.
\end{equation}

\item{(b)} it holds that
\begin{equation}
\label{new ax: (b) alpha region cond} \alpha(q) \in
\begin{dcases}
(-\infty,0] 
\;\; &\hbox{for} \;\; \phi(q) > 0 \\
\quad
[0 , 1] \;\; &\hbox{for} \;\; \phi(q) < 0.
\end{dcases}
\end{equation}

\end{description}

\end{theorem}

\begin{remark}
The axiom [gsk4] reduces to [GSK4] if we choose $\alpha(q)=1-q$.
\end{remark}

\begin{remark}
The conditions for $\alpha(q)$ and $\phi(q)$ are identical to the
conditions from subsection \ref{entropy as expected ic}.
Accordingly, the class of entropy functionals (\ref{suyari: Sq_g}) is the same as
the class (\ref{invar: S unnor}).
\end{remark}

\begin{IEEEproof}
By straightforward repetition of steps from  to Suyari's proof,
the form (\ref{suyari: Sq_g}) is uniquely determined by solving
functional equation (\ref{q-additivity_g}) from [gsk4] and by using
the continuity with respect to $p$ as assumed in [GSK1].

The properties of $\phi(q)$ given by (a) straightforwardly follow from continuity of $S_q$ and $\alpha(q)$. Before proving the equality (\ref{new ax: alpa phi limit}), we make the following note. The condition (\ref{new ax: alpa phi limit}) is necessary and sufficient
%
for satisfaction of the limit property (\ref{Shannonentropy_lim}),
\begin{equation}
\label{new ax: sum lim = sum condition}
\qlim{1} {1 - \sum_{i=1}^n p_i^{-\alpha(q) + 1} \over
\phi(q)}=%
-\sum_{i=1}^n k \cdot p_i \ln p_i.
\end{equation}
To prove this,
%
%
let us introduce $\gamma(q)= p^{- \alpha(q)} - 1$. Using $\gamma(q) \ra 0$ when $q
\rightarrow 1$ and $(1 + t )^{1 \over t} \ra e$ when $t
\rightarrow 0$, we have
\begin{align}
\label{new axx: qlim 1 -p / phi}
\qlim{1}{1 - p^{-\alpha(q)} \over \phi(q)} &=%
\qlim{1}{\alpha(q) \over \phi(q)}\cdot%
{1 - p^{-\alpha(q)} \over\alpha(q)} =\nonumber\\%
=& \qlim{1} {\alpha(q) \over \phi(q)} \cdot  {\ln p \over \ln
\left(1 + \gamma(q)\right)^{1 \over \gamma(q) }}\nonumber\\%
=& \qlim{1} {\alpha(q) \over \phi(q)} \cdot  \ln p.
\end{align}
%
For $p_i = p = 1/n$ (\ref{new ax: sum lim = sum condition}) reduces to
\begin{equation}
\qlim{1}{1 - p^{-\alpha(q)} \over \phi(q)} %
= - k \cdot  \ln p,
\end{equation}
and according to (\ref{new axx: qlim 1 -p / phi}), the condition (\ref{new ax: alpa phi limit}) is necessary. On the other hand,
\begin{equation}
\qlim{1} {1 - \sum_{i=1}^n p_i^{-\alpha(q) + 1} \over
\phi(q)}=%
\qlim{1} \sum_{i=1}^n p_i \cdot {1 - p_i^{-\alpha(q)} \over
\phi(q)}
\end{equation}
and (\ref{new axx: qlim 1 -p / phi}) impiles the sufficiency of condition (\ref{new ax: alpa phi limit}).

%
%

Property (b) can be proven by taking the second derivative of
${I_q\left( p \right)}$ with respect to $p$, which should be
nonnegative for any fixed $q\in \sR^+$, since $I_q(p)$ is convex
by [T2]. Thus, we can derive a constraint
\begin{equation}
\label{alpha constraint general}%
{\alpha(q) \over \phi(q)}\cdot\left(\alpha(q) -1
\right)\ge 0
\end{equation}
for any $q\in \sR^+$. The constraint
(\ref{alpha constraint general}) is satisfied if
\begin{align}
\label{alpha constraint cases general} \alpha(q) \in
\begin{dcases}
(-\infty,0] \cup [1, \infty) \;\; &\hbox{for} \;\; \phi(q) > 0 \\
\quad\quad\quad [0 , 1] \;\; &\hbox{for} \;\; \phi(q) < 0.
\end{dcases}
\end{align}
As shown in \cite{ilic2012entropy}, if $\alpha(q)$ and $\phi(q)$ are continuous
and the equality (\ref{new ax: alpa phi limit}) holds, then
$\alpha(q) \not\in [1, \infty)$, and the equality (\ref{new ax:
(b) alpha region cond}) follows, which proves and theorem.

\end{IEEEproof}

\section{Conclusion}
In this paper, we reviewed generalized Shannon-Khinchin axioms proposed by Suyari in \cite{suyari2004generalization}. 

We discussed Suyari's proof of a unique class of functions satisfying those axioms, pointed out the oversight, supported it with a counterexample and gave the correction of the proof. Suyari's paper has been widely cited and a similar oversight has been noticed in some of them. For example, in \cite{furuichi2005uniqueness} the author follows the same procedure in generalizations of Hobson's axioms, which leads to the similar incorrectness.  

In addition, we generalize Suyari's axioms characterizing the recently introduced class of
entropies obtained by averaging pseudoadditive information content
introduced in \cite{ilic2012entropy}.


%
%

\bibliographystyle{IEEEtran}
\bibliography{reference}

\end{document}